\documentclass[sigconf,nonacm]{acmart}

\setcopyright{none}
\renewcommand\footnotetextcopyrightpermission[1]{}
\newcommand{\system}{\texttt{PrismaDV}}

\newcommand{\header}[1]{\vspace{1mm}\noindent\textbf{#1}.}

\newcommand{\headerul}[1]{\vspace{1mm}\noindent\underline{\textit{#1}}.}

\usepackage{enumitem}
\usepackage{tikz}
\usepackage{cleveref}
\DeclareRobustCommand*\circled[1]{\tikz[baseline=(char.base)]{\node[shape=circle,fill=black,inner sep=1pt,font=\small\bfseries\color{white}](char){#1};}}

\begin{document}

\title{``Will This Data Break My Task?'' Interactive~Synthesis~of~Task-Aware~Data~Unit~Tests}

\author{Hao Chen}
\orcid{0009-0004-1887-6630}
\affiliation{%
  \institution{BIFOLD \& TU Berlin}
  \country{}
}
\email{hao.chen@tu-berlin.de}

\author{Arnab Phani}
\orcid{0009-0001-2935-0608}
\affiliation{%
  \institution{BIFOLD \& TU Berlin}
  \country{}
}
\email{arnab.phani@tu-berlin.de}

\author{Sebastian Schelter}
\orcid{0000-0003-4722-5840}
\affiliation{%
  \institution{BIFOLD \& TU Berlin}
  \country{}
}
\email{schelter@tu-berlin.de}

% !TEX root = ../main.tex
\begin{abstract}

Data is a central resource for modern enterprises and institutions, and data errors propagating through data pipelines lead to serious impact in production. Therefore, data validation is essential for ensuring the reliability of downstream applications. This led to the development of data unit tests, executable programs that test data before moving it around through large data pipelines. However, existing frameworks derive data unit tests from observed data alone, ignoring the semantics of the code that consumes the data downstream. 

To this end, we present \system{}, a compound AI system that synthesizes task-aware data unit tests for tabular data by jointly analyzing data and downstream task code. \system{} decomposes the test generation into multiple LLM-powered steps: data profiling, detection of column accesses, data flow analysis in the task code, and the inference of implicit data assumptions. It subsequently synthesizes code for the data unit test, and maintains an internal ``data-code assumption graph'' that links generated data constraints back to the task's source code. 

We demonstrate \system{} through an interactive web-based interface where attendees run the system on five real-world datasets with 60 downstream tasks, synthesize, inspect and refine both natural language assumptions about the data and executable data constraints. The interface allows attendees to navigate the data-code assumption graph, compare task-aware data unit tests against task-agnostic baselines on erroneous data batches, and interactively edit assumptions and data constraints. Furthermore, attendees can observe how a custom prompt optimizer adapts the system to specific datasets over time.
\end{abstract}

\maketitle

% !TEX root = ../main.tex
\section{Introduction}

Data is a central resource for modern enterprises and institutions, and data issues, such as missing or incorrect information, can seriously impact their operations. Data errors propagating through data pipelines cause serious impact in production, such as outages of mobile apps~\cite{facebookios} or the loss of medical records~\cite{ukcorona}. Furthermore, data errors often lead to the silent performance degradation of deployed ML models~\cite{10.1145/3500923,polyzotis2019data,nigenda2022amazon}. A reason for this is that many organizations have adopted a ``collect first, analyze later'' workflow, relying on the schema-on-read interpretation of data in downstream applications. Therefore, corrupted data often propagates unnoticed and causes havoc in production.

\header{Shortcomings of current data unit testing frameworks} As a consequence, data unit testing frameworks such as TensorFlow Data Validation~\cite{polyzotis2019data} (TFDV), Amazon's Deequ~\cite{schelter2018automating}, and Great Expectations~\cite{greatexpectations} have become widely used for tabular data in industry in recent years. These frameworks enable engineers to define data unit tests via declarative data constraints that will be evaluated against unseen data batches. Major cloud providers offer data unit testing as part of their data infrastructure, e.g., via AWS Glue Data Quality~\cite{awsglue}, Databricks Pipeline Expectations~\cite{databricksexpectations}, or Google~Dataplex~\cite{googledataplex}.
\begin{figure*}[t]
  \centering
  \includegraphics[width=\textwidth]{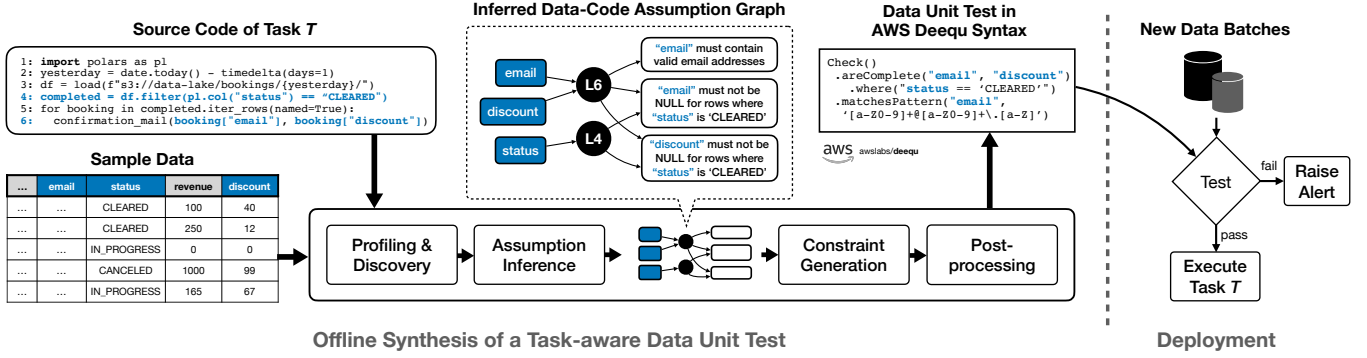}
  \caption{Overview of \system{}. Given the source code of a downstream task and a data sample, the \system{} system synthesizes executable data unit tests in Deequ or Great Expectations syntax. For that, it performs data profiling and data flow analysis, assumption inference via the data-code assumption graph, constraint code generation, and post-processing. The system infers implicit data assumptions in natural language, linked to specific code lines. At deployment time, the generated data unit test is evaluated on each new data batch; passing batches are forwarded to the downstream task, while failing batches trigger an alert.}
  \label{fig:system}
\end{figure*}
Despite their popularity, existing frameworks suffer from severe shortcomings. Authoring and maintaining data unit tests remains tedious and error-prone, especially for tables with hundreds of columns. Frameworks such as Deequ and TFDV alleviate this burden by automatically suggesting constraints from sample data, but the resulting data unit tests still require manual post-editing by a data engineer with domain knowledge. This is because heuristically suggested constraints are often either too strict or too general; overly strict tests produce false alarms, leading to alert fatigue~\cite{10.1145/3583780.3614786}, while overly general tests miss domain-specific data errors. 

\header{The case for task-aware data unit test generation} A major limitation of current approaches is that they {\em rely on observed data only and ignore the characteristics of the downstream tasks that consume the data}~\cite{chen2025tadv,chen2026prismadv}. This leads to several missed opportunities. First, certain downstream tasks might only access parts of the data  (especially for large denormalized datasets common in enterprise data lakes), which means that data unit tests for these tasks should focus on the accessed columns only. Second, the code of downstream tasks is often written by experienced data engineers, with implicit domain knowledge about the data ``baked in'', which may be helpful to extract into a data unit test. Therefore, the data unit tests should be {\em specialized to the downstream tasks} for which they are deployed. However, this specialization is inherently difficult as it requires an ``understanding'' of downstream task code. In practice, production data pipelines support a diverse set of downstream tasks, ranging from recurring BI/ETL processing to web applications, or feature engineering and ML training or inference, and these tasks often encode different semantics and assumptions about the same data~\cite{rezig2021datacleaning}.

\header{Synthesis of task-aware data unit tests with \system{}} To this end, we demonstrate \system{}, a task-aware data validation system that interactively synthesizes specialized data unit tests for individual downstream tasks by jointly analyzing data and task code~(\Cref{sec:system}). \system{} is a compound AI system~\cite{compound-ai-blog}, which decomposes task-aware data unit test generation into multiple steps: data profiling, detection of accessed columns, dataflow analysis in the task code, and the inference of implicit data assumptions (e.g., expected value ranges, non-null columns or conditional dependencies) in natural language, followed by the generation of data constraint code for the data unit test.  The system leverages the code understanding and code synthesis capabilities of LLMs, and maintains a ``data-code assumption graph'' that traces each synthesized data constraint back to the inferred data assumption that justifies it and to the corresponding lines in the task source code. This decomposition enables \system{} to generate targeted, task-specific data unit tests while providing full provenance from the data unit test back to the validated data and downstream code.

\header{Demonstration details} We demonstrate \system{} with an interactive web interface, together with a collection of five real-world datasets accompanied by 60~downstream tasks~(\Cref{sec:demo}). Attendees run the full test generation pipeline end-to-end, inspect and refine intermediate outputs through the data-code assumption graph, and compare the generated task-aware tests against task-agnostic baselines. We additionally demonstrate how \system{} adapts to specific datasets over time via a custom prompt optimizer based on collected task outcomes. We provide the source code, datasets, and the web-based interface at \textcolor{blue}{\url{https://github.com/deem-data/prismadv-demo}}.

% !TEX root = ../main.tex
\section{System Overview}
\label{sec:system}

We introduce task-aware data validation and the \system{} system (\Cref{fig:system}), see our full paper~\cite{chen2026prismadv} for further details and experiments.

\header{Task-aware data unit tests} Consider a downstream task~$T$, that consumes tabular data over time in batches, (e.g., a Python program computing a daily report over the latest customer orders in an e-commerce platform). The task's source code accesses certain parts of the data and contains implicit assumptions about the data, such as expected value ranges, non-null columns, or inter-column dependencies. The goal of task-aware data unit test generation is to automatically synthesize a data unit test (in the form of executable data constraints) from~$T$ and a data sample. Each constraint implements a boolean function to validate a certain assumption about the data (e.g., by comparing an aggregate statistic computed over the data to a fixed threshold). These constraints will be evaluated against each new unseen data batch before it is forwarded to~$T$; if all constraints pass, the batch is considered safe for~$T$ to consume, if the test fails, the data batch is quarantined and engineers are alerted to manually intervene and fix the data.

\begin{figure*}[t!]
  \centering
  \includegraphics[width=\linewidth]{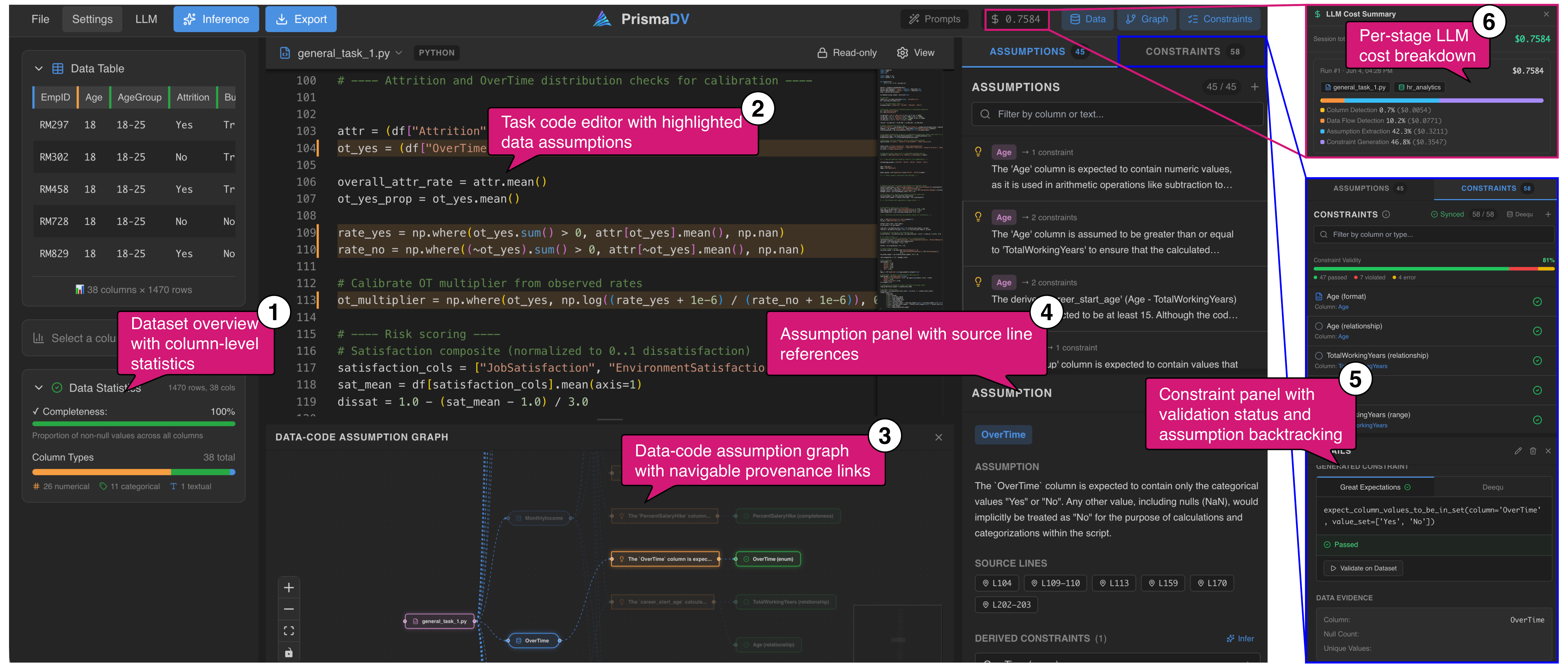}
  \caption{\system{} web interface. \circled{1}~Dataset overview with data table preview, column statistics (count, null rate, unique values, value distribution), and overall data quality metrics. \circled{2}~Task code editor with syntax highlighting and data flow annotations; yellow-highlighted lines indicate where the task accesses or transforms columns. \circled{3}~Data-code assumption graph showing the bipartite connections between the task code, accessed columns, and inferred assumptions; clicking a node navigates to the corresponding artifact. \circled{4}~Assumption panel listing each assumption with its natural-language description, source line references, and derived constraints. \circled{5}~Constraint panel showing generated data constraints in Great Expectations and Deequ syntax, with their validation status. \circled{6}~Per-stage LLM cost breakdown with per-run token usage across column detection, data flow analysis, assumption extraction, and constraint generation.}
  \label{fig:ui}
  \vspace{0.35cm}
\end{figure*}

\header{Architecture and implementation of \system{}} \system{} is a compound AI system that decomposes task-aware data unit test generation into a multi-step pipeline, as shown in \Cref{fig:system}. Given task code~$T$ and a data sample, the system performs data profiling, column access detection, data flow detection, assumption inference, and constraint code generation, where each LLM-powered step progressively builds on the outputs of the previous ones. The results are organized in the ``data-code assumption graph'' that maintains the full provenance from generated data constraints back to the source code of the task. The LLM-powered components are implemented via DSPy~\cite{khattab2023dspy}.

\headerul{Step~1: Data profiling \& Column access detection} \system{} first computes a lightweight data profile over the data sample, which is used as context for the subsequent stages. Next, \system{} determines the columns that are actually accessed by the downstream task. The column access detection module receives the task code, a textual description of the task, and the available dataset columns and their profiles, and prompts an LLM to identify which columns the task operates on. The module returns the subset of columns that the code reads, filters, transforms, or otherwise depends on. This design is motivated by prior work~\cite{chen2025tadv}, which shows that an LLM reliably detects accessed columns even for non-trivial cases such as SQL \texttt{SELECT *} queries with \texttt{EXCLUDE} clauses, where purely static or string-matching analysis falls short.

\headerul{Step~2: Construction of the data-code assumption graph} Next, assumption extraction is performed in parallel over accessed columns. For each column, \system{} first analyzes the data flow in the code and annotates the corresponding line numbers that operate on the column or data derived from it. Finally, the LLM infers a list of assumption objects, each containing: (i)~a natural-language description of the implicit data assumption, (ii)~the involved column, and (iii)~the source lines in~$T$ that motivated the inference.

\headerul{Step~3: Constraint code synthesis} The final generation stage synthesizes executable constraints from the inferred assumptions, targeting two validation frameworks: Great Expectations and Deequ. The LLM receives each assumption together with the task code, the downstream task description, the accessed columns, and a summary of the available API functions in Deequ and Great Expectations.  \system{} then synthesizes data constraint code; together with the preceding steps, this produces the complete data-code assumption graph, connecting accessed columns via code lines to assumptions, which are the basis for executable data constraints, as illustrated in \Cref{fig:system}.

\headerul{Step~4: Postprocessing} Finally, \system{} filters out constraints that cannot be successfully parsed by Deequ or Great Expectations, as well as constraints that do not pass on the data sample at hand. 

\header{Detection quality} To exemplify the benefits of \system{}, we summarize an experiment from \cite{chen2026prismadv} on a custom benchmark with five datasets, 60 downstream tasks, and 125 data batches to test, obtained by injecting synthetic errors spanning structural, integrity, numerical, textual, and format corruptions into clean data samples. Each method has to predict whether a data batch is safe to forward to a downstream task or causes a crash or erroneous output. \Cref{tab:eidbench} shows that \system{} substantially outperforms both task-agnostic baselines (\texttt{deequ}, \texttt{tensorflow-dv}) and LLM-based prompting and agentic baselines~\cite{yang2024swe}, improving the F1 score by more than 20~points over the strongest competitor.

\begin{table}[b!]
\centering
\caption{Detection quality for potentially erroneous data batches for downstream tasks. Best F1 score in bold, second best underlined. \system{} outperforms all baselines by a large margin of more than 20~points.}
\label{tab:eidbench}
\small
\setlength{\tabcolsep}{4pt}
\renewcommand{\arraystretch}{0.95}
\begin{tabular}{@{}lccc@{}}
\toprule
\textbf{Method} & \textbf{Precision} & \textbf{Recall} & \textbf{F1} \\
\midrule
\texttt{deequ}                        & 65.8\% & 14.4\% & 24.2\% \\
\texttt{tensorflow-dv}                & 65.1\% & 42.7\% & 50.3\% \\
\midrule
\texttt{zero-shot [gemini-2.5-pro]} & 65.0\% & 22.8\% & 31.0\% \\
\texttt{zero-shot [gpt-5]}          & 80.3\% &  8.7\% & 31.9\% \\
\texttt{few-shot [gemini-2.5-pro]}    & 69.9\% & 26.8\% & 43.0\% \\
\texttt{few-shot [gpt-5]}             & 75.8\% & 21.4\% & 47.2\% \\
\texttt{swe-agent [gpt-5]}            & 78.6\% & 20.3\% & 47.2\% \\
\midrule
\system{}~\texttt{[gemini-2.5-pro]}   & 76.1\% & 73.0\% & \underline{73.9\%} \\
\system{}~\texttt{[gpt-5]}            & 70.5\% & 86.6\% & \textbf{77.4\%} \\
\bottomrule
\end{tabular}
\end{table}

\header{Optimizing \system{} from execution feedback} As data unit tests run in production, engineers collect task and test outcomes over time, and can use a custom prompt optimizer in \system{} to refine the prompts for its LLM-based modules. This reduces the need for manual tuning and improves the quality of the data unit tests over time. In these collected outcomes, passing data constraints provide little signal, therefore the optimizer of \system{} focuses on constraint failures that coincide with task success, which indicate false alarms. It backtraces these cases through the data-code assumption graph to identify problematic assumptions and code patterns, and uses them as context for prompt optimization. See our full paper \cite{chen2026prismadv} for details.

% !TEX root = ../main.tex
\section{Demonstration Details}
\label{sec:demo}

We demonstrate \system{} through an interactive web interface, shown in \Cref{fig:ui}, which provides access to all intermediate outputs and supports interactive refinement of assumptions and constraints.

\header{Overview} We organize the demonstration in three stages. 

\begin{enumerate}[leftmargin=*]
  \item First, attendees run the fully automated test generation pipeline on a live LLM backend, where \system{} synthesizes task-aware data unit tests end-to-end. They explore the generated tests through the data-code assumption graph and compare the results against task-agnostic baselines~(\Cref{sec:auto}).
  \item Second, attendees interactively refine the generated assumptions and constraints through the web interface, re-generate the resulting tests and observe how changes propagate through the data-code assumption graph~(\Cref{sec:interactive}).
  \item Third, we demonstrate how \system{} adapts its prompts over time using task and test outcomes from deployed data unit tests. Here, attendees run our custom prompt optimizer and inspect the resulting changes as a side-by-side diff~(\Cref{sec:adaptation}).
\end{enumerate}

\subsection{Synthesis of Task-Aware Data Unit Tests}
\label{sec:auto}
We first walk attendees through the fully automated pipeline, where \system{} takes a dataset and task code as input and progressively constructs the data-code assumption graph, and produces an executable data unit test.

\header{Data and Tasks} We provide five datasets spanning diverse distributions of numerical, categorical, and textual columns, together with 60 executable tasks, all taken from a benchmark proposed in~\cite{chen2026prismadv}; the tasks were created via an LLM-assisted, human-in-the-loop pipeline and carefully manually reviewed and post-edited to ensure correctness and consistency. The tasks span a broad range of downstream application types, from simple aggregation queries to web applications to ML feature engineering, with implicit data assumptions of varying complexity embedded in the code. Attendees can also upload their own datasets and task code. After selecting  a task, attendees review the data statistics and task code, and configure their preferred language model.

\header{Generation} Next, attendees trigger the test generation pipeline, which first runs column access detection to identify which columns the task code depends on. Attendees then choose to validate all accessed columns or a selected subset. The system proceeds with data flow analysis to locate the specific code lines operating on each column, followed by assumption inference and constraint generation. The system runs against a live LLM backend, to showcase actual generation latencies and API costs.

\header{Exploration} Attendees inspect the generated artifacts through the data-code assumption graph~(\circled{3} in \Cref{fig:ui}). Clicking on an assumption navigates the code editor~(\circled{2}) to the relevant source lines; hovering reveals the assumption text in the assumption panel~(\circled{4}). Each assumption links to its derived constraints, and each constraint can be backtracked to its underlying assumptions and accessed columns in the constraint panel~(\circled{5}). A per-stage cost breakdown~(\circled{6}) shows the LLM usage across column detection, assumption extraction, and constraint generation. The system automatically validates each constraint against the current dataset, reporting pass/fail status and flagging violating rows.

\subsection{Interactive Refinement}
\label{sec:interactive}

While the automated pipeline produces a high-quality initial set of assumptions and constraints, domain experts can continue to polish it by incorporating business rules not expressed in the code, or adjusting constraints to specific deployment contexts. The interactive refinement capabilities of \system{} allow them to use the generated tests as a starting point and efficiently arrive at a customized, production-ready test. Crucially, our system simplifies this process: attendees need to review and refine assumptions only for the columns a task actually accesses, rather than the full schema. Refinement is moreover optional, as the automated pipeline already produces a working data unit test, so experts only need to intervene on the data constraints they disagree with.

\header{Refining assumptions} Attendees review the generated assumptions in the assumption panel~(\circled{4}) and can remove any assumption which they find incorrect or irrelevant; the corresponding constraints are automatically deleted. If an assumption is close to the intended requirement but not accurate, attendees can edit its text and trigger the constraint generation module to produce updated constraints based on the revised assumption. Attendees can also add entirely new assumptions that the system missed and trigger inference to generate the corresponding data constraints. When an assumption already has linked constraints, \system{} runs a gap analysis to determine whether the existing constraints sufficiently cover the assumption and only generates the missing constraints. For instance, an attendee may notice that the system inferred ``status must be one of CLEARED, IN\_PROGRESS, CANCELLED'' but the actual domain includes a fourth value ``PENDING''. Editing the natural language assumption triggers re-generation, and the updated constraint will then include all four values.

\header{Refining constraints} In the constraint panel~(\circled{5}), attendees can directly edit or delete individual constraints. After editing a constraint, they can validate its syntax against the target framework and test whether it passes or fails on the current data sample. Attendees can also add new constraints and link them to an existing assumption.
The interface provides immediate visual feedback as assumptions are modified and constraints are regenerated, allowing attendees to observe how changes propagate through the data-code assumption graph.

\subsection{Domain Adaptation}
\label{sec:adaptation}

Finally, we showcase how the custom prompt optimizer in \system{} improves its module prompts and the resulting data unit tests, based on historical task and test outcomes.

\header{Demonstration workflow} We provide pre-collected execution histories for each dataset, containing pass/fail outcomes across multiple data batches and downstream tasks. Attendees trigger the optimization and inspect the resulting prompt changes as a side-by-side diff, showing how the optimizer adapted the instructions to reduce false alarms and improve constraint quality for the specific dataset. They can then re-run the generation pipeline with the optimized prompts and compare the updated constraints against the previous version.

\bibliographystyle{ACM-Reference-Format}
\bibliography{references}

\end{document}